\documentclass[pra,showpacs,preprintnumbers,amsmath,amssymb,twocolumn]{revtex4}

\usepackage{amssymb}
\usepackage{amsmath}
\usepackage{graphicx}
\usepackage{epsfig}
\usepackage{subfigure}
\usepackage{amsfonts}
\usepackage{CJK}

\begin{document}

\title{Proposal for realizing a multiqubit tunable phase gate of one qubit
simultaneously controlling $n$ target qubits using cavity QED}

\author{Chui-Ping Yang$^{1,2}$, Qi-Ping Su$^{1}$, and Jin-Ming Liu$^{2}$}

\address{$^1$Department of Physics, Hangzhou Normal University,
Hangzhou, Zhejiang 310036, China}

\address{$^2$State Key Laboratory of Precision Spectroscopy,
Department of Physics, East China Normal University, Shanghai
200062, China}

\date{\today}

\begin{abstract}
We propose a way to realize a multiqubit tunable phase gate of one
qubit simultaneously controlling $n$ target qubits with atoms in
cavity QED. In this proposal, classical pulses interact with atoms
{\it outside} a cavity only, thus the experimental challenge of
applying a pulse to an intra-cavity single atom without affecting
other atoms in the same cavity is avoided. Because of employing a
first-order large detuning, the gate can be performed fast when
compared with the use of a second-order large detuning.
Furthermore, the gate operation time is independent of the number
of qubits. This proposal is quite general, which can be applied
to various superconducting qubits coupled to a resonator, NV centers
coupled to a microsphere cavity or quantum dots in cavity QED.

\end{abstract}

\pacs{03.67.Lx, 42.50.Dv} \maketitle
\date{\today}

\begin{figure}[tbp]
\includegraphics[bb=95 461 525 688, width=9.0 cm, height=4.5 cm, clip]{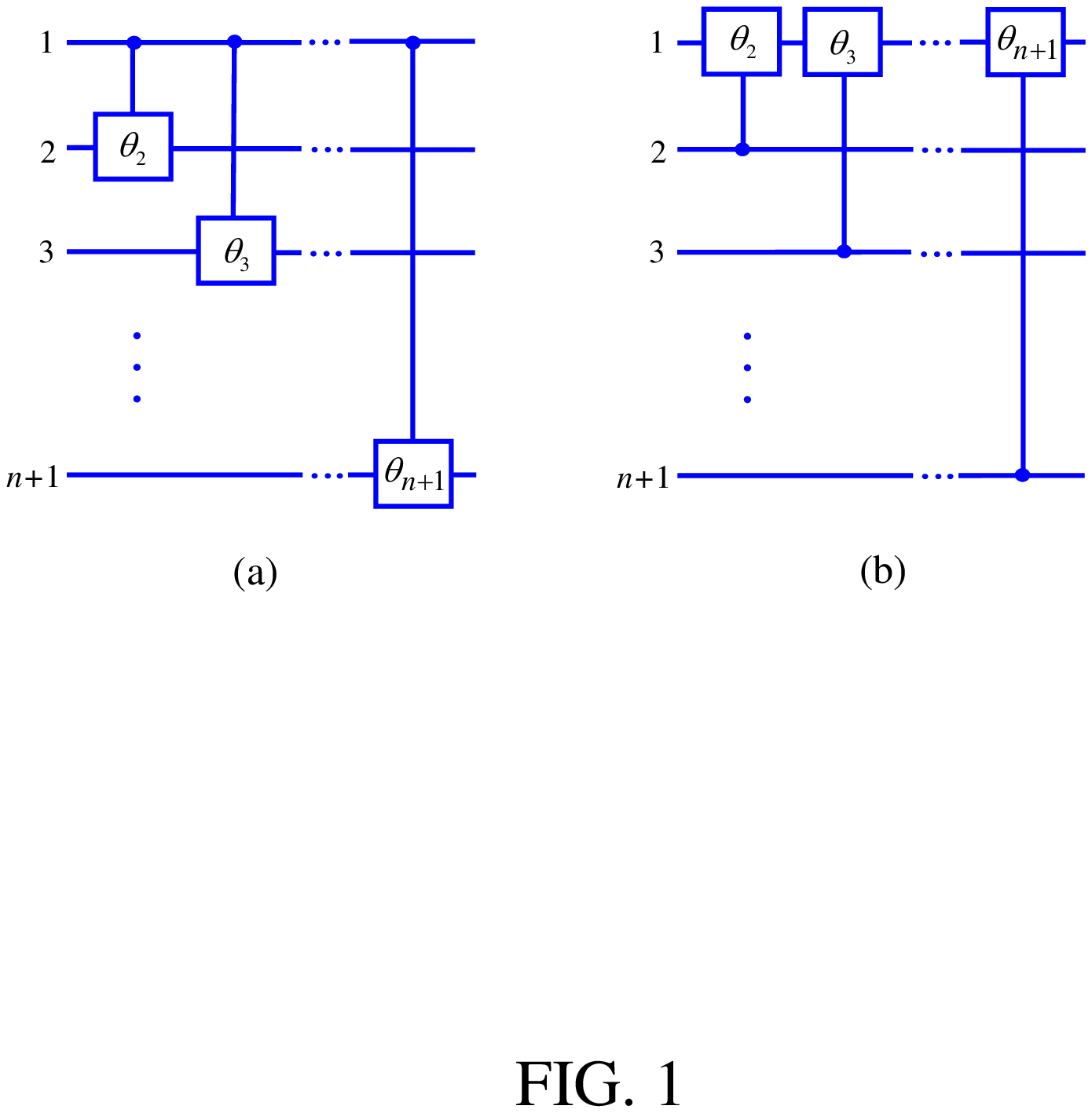} %
\vspace*{-0.08in}
\caption{(Color online) (a) Schematic circuit of a multiqubit controlled-phase
(CP) gate with qubit 1 (on top) \textit{simultaneously} controlling $n$
target qubits ($2,3,...,n+1$). (b) The circuit for a phase gate with many control
qubits ($2,3,...,n+1$) acting on one target (qubit 1).}
\label{fig:1}
\end{figure}

\textit{Introduction.}---Over the past decade, many proposals have been
presented for realizing a multiqubit controlled-phase (CP) or controlled-NOT
gate with \textit{multiple-control} qubits acting on \textit{one target}
qubit, using various physical systems [1-7]. These proposals are important
because they opened new avenues for realizing this type of multiquibit
controlled gates, which are of significance in quantum information
processing (QIP).

In this work we focus on a multiqubit tunable phase gate of one qubit
simultaneously controlling $n$ target qubits. This gate is useful in QIP
(e.g., it has an important application in quantum Fourier transforms). In
the following, we will present a way for implementing this gate with atoms
in cavity QED.

The proposal has these features: (i) The atoms interact with classical
pulses \textit{outside} the cavity, thus the experimental challenge of
applying a pulse to an intra-cavity single atom without affecting (many)
other atoms in the same cavity is avoided; (ii) The gate operation employs a
first-order large detuning, thus the gate can be performed fast when
compared with the use of a second-order large detuning; and (iii) The
operation time is independent of the number of qubits and thus does not
increase with the number of qubits.

\textit{Multiqubit tunable phase gate.}---The multiqubit phase gate here
consists of $n$ two-qubit CP gates as depicted in Fig.~1(a). Each two-qubit
CP gate has a shared control qubit (qubit $1$) but a \textit{different}
target qubit ($2,3,...,$ or $n+1$). This multiqubit phase gate has the
property: (i) when the control qubit $1$ is in the state $\left|
1\right\rangle ,$ phase shifts $e^{i\theta _2},$ $e^{i\theta _3},...,$ and $%
e^{i\theta _{n+1}}$ are simultaneously induced to the state $\left|
1\right\rangle $ of the target qubits $2,$ $3,...,$ and $n+1$ respectively,
but nothing happens to the state $\left| 0\right\rangle $ of each target
qubit; (ii) when the control qubit $1$ is in the state $\left|
0\right\rangle ,$ both states $\left| 0\right\rangle $ and $\left|
1\right\rangle $ of each target qubit remain unchanged. Here, $\theta
_2,\theta _3,...,$and $\theta _{n+1}$ are adjustable as described below,
taking values from $0$ to $2\pi .$

\textit{Atom-cavity dispersive interaction.}---Consider atoms ($2,3,...,n+1$%
) with four levels depicted in Fig.~2(a). The cavity mode is coupled to the $%
\left| 2\right\rangle \leftrightarrow \left| 3\right\rangle $ transition of
each atom but highly detuned (decoupled) from the transition between any
other two levels [Fig.~2(a)]. In the interaction picture, the interaction
Hamiltonian is given by

\begin{equation}
H=\hbar \sum_{k=2}^{n+1}g(e^{-i\Delta _ct}a^{+}\sigma _{23,k}^{-}+\text{H.c.}%
),
\end{equation}
where $\Delta _c=\omega _{32}-\omega _c$ is the detuning of the cavity
frequency $\omega _c$ with the $\left| 2\right\rangle \leftrightarrow \left|
3\right\rangle $ transition frequency $\omega _{32}$ of the atoms, $g$ is
the coupling constant between the cavity mode and the $\left| 2\right\rangle
\leftrightarrow \left| 3\right\rangle $ transition, and $\sigma
_{23,k}^{-}=\left| 2\right\rangle _k\left\langle 3\right| .$

\begin{figure}[tbp]
\includegraphics[bb=178 377 397 604, width=7.0 cm, height=5.5 cm, clip]{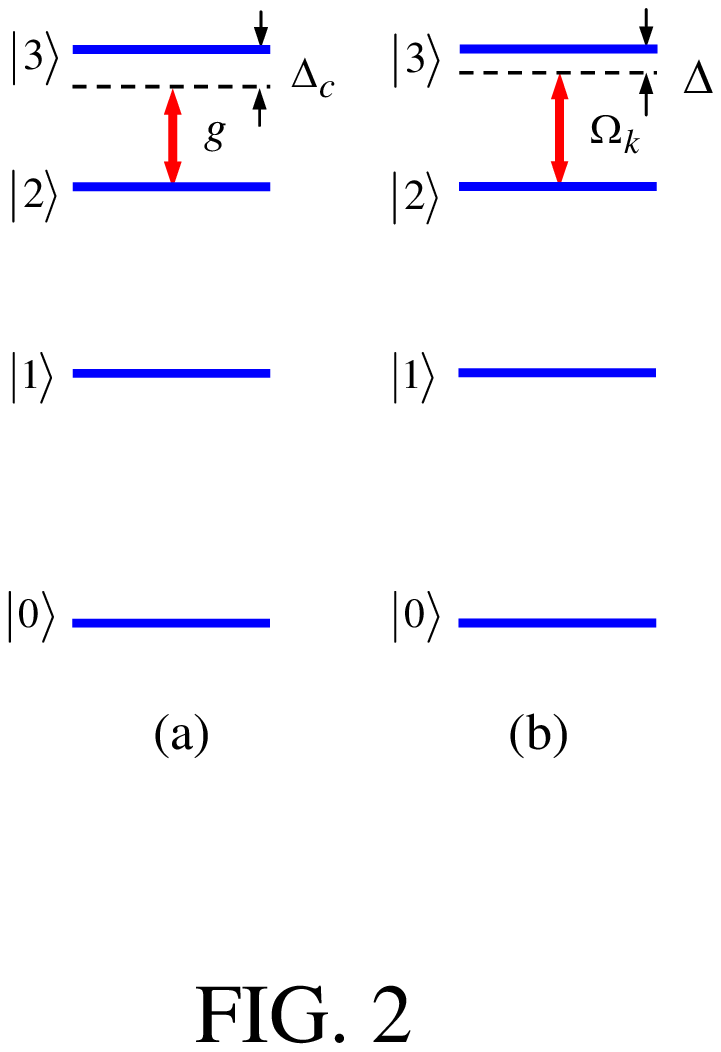} %
\vspace*{-0.08in}
\caption{(Color online) (a) Atom-cavity off-resonant interaction. (b)
Atom-pulse off-resonant interaction.}
\label{fig:2}
\end{figure}

For $\Delta _c\gg g$ (i.e., the cavity mode is dispersively coupled to the
$\left| 2\right\rangle \leftrightarrow \left| 3\right\rangle $ transition of
each atom), no energy exchange occurs between the atoms and the
cavity mode. In this case, when the level $\left| 3\right\rangle $ of each
atom is not excited, the Hamiltonian~(1) becomes [8-10]
\begin{equation}
H_{\mathrm{eff}}=-\hbar \sum_{k=2}^{n+1}\frac{g^2}{\Delta _c}a^{+}a\sigma
_{22,k}.
\end{equation}

The time-evolution operator corresponding to $H_{\mathrm{eff}}$ is given by $%
U(t)=\otimes _{k=2}^{n+1}U_{kc}\left( t\right) ,$ where $U_{kc}\left(
t\right) =\exp [i\left( g^2/\Delta _c\right) a^{+}a\sigma _{22,k}t]$ acts on
the cavity mode and atom $k$ ($k=2,3,...,n+1$). It is easy to verify that
for $t=\pi \Delta _c/g^2,$ $U_{kc}\left( t\right) $ leads to the
transformation $\left| 2\right\rangle _k\left| 1\right\rangle _c\rightarrow
-\left| 2\right\rangle _k\left| 1\right\rangle _c,$ while leaves the states $%
\left| 0\right\rangle _k\left| 0\right\rangle _c,$ $\left| 1\right\rangle
_k\left| 0\right\rangle _c,$ $\left| 2\right\rangle _k\left| 0\right\rangle
_c,$ $\left| 0\right\rangle _k\left| 1\right\rangle _c$ and $\left|
1\right\rangle _k\left| 1\right\rangle _c$ unchanged$.$ Here, $\left|
1\right\rangle _c$ is the single-photon state of the cavity.

\textit{Atom-pulse dispersive interaction}---Consider a four-level atom $k$,
which is driven by a pulse with frequency $\omega $ and initial phase $\phi
=0.$ The pulse is coupled to the $\left| 2\right\rangle \leftrightarrow
\left| 3\right\rangle $ transition of atom $k$ with a detuning $\Delta
=\omega _{32}-\omega $ [Fig. 2(b)], which is described by

\begin{equation}
H_I^k=\hbar \left( \Omega _ke^{-i\Delta t}\left| 2\right\rangle \left\langle
3\right| +\text{H.c.}\right) ,
\end{equation}
where $\Omega _k$ is the pulse Rabi frequency. For $\Delta \gg \Omega _k$,
the Hamiltonian~(3) becomes $H_I^k=\hbar \frac{\Omega _k^2}\Delta \left(
\left| 3\right\rangle \left\langle 3\right| -\left| 2\right\rangle
\left\langle 2\right| \right) ,$ under which a pulse of duration $t_k$ leads
to $\left| 2\right\rangle \rightarrow e^{i\theta _k}\left| 2\right\rangle $
with $\theta _k=\Omega _k^2t_k/\Delta .$ The $\theta _k$ is adjustable by
changing the Rabi frequency $\Omega _k$ or the duration $t_k$ of the pulse.

\textit{Gate implementation}---Consider a two-level atom $1$ with two levels
$\left| 0\right\rangle $ (ground) and $\left| 1\right\rangle $ (excited),
and $n$ identical atoms ($2,3,...,n+1$) with four levels depicted in Fig. 2.
For each atom, the two lowest levels $\left| 0\right\rangle $ and $\left|
1\right\rangle $ represent the two logical states of a qubit.

Our gate operation needs to move atoms into or out of the cavity. This can
be achieved by employing a translating optical lattice trap [11,12].
Experimentally, using an optical lattice trap to deliver atoms or a single
atom into an optical microcavity has been reported [12], and by controlling
the motion of the standing wave, a single atom can be transported to a
preselected point along the standing wave with a very high precision [11].

We suppose that during the gate operation, (i) the cavity mode is resonant
with the $\left| 0\right\rangle \rightarrow $ $\left| 1\right\rangle $
transition of atom $1$ with a coupling constant $g_r;$ and (ii) the cavity
mode is dispersively coupled to the $\left| 2\right\rangle \leftrightarrow
\left| 3\right\rangle $ transition of atoms ($2,3,...,n+1$) but highly
detuned (decoupled) from the transition between any other two levels of
atoms ($2,3,...,n+1$). In addition, the following gate operation involves
the application of a classical pulse resonant to the $\left| 1\right\rangle
\leftrightarrow \left| 2\right\rangle $ transition (with frequency $\omega
_{21}$) of atoms ($2,3,...,n+1$). The pulse Rabi frequency is denoted as $%
\Omega _r.$ The frequency, initial phase, and duration of the resonant pulse
are denoted as \{$\omega ,$ $\phi ,$ $t^{\prime }$\} below.

\begin{figure}[tbp]
\includegraphics[bb=44 174 518 690, width=8.0 cm, height=5.5 cm, clip]{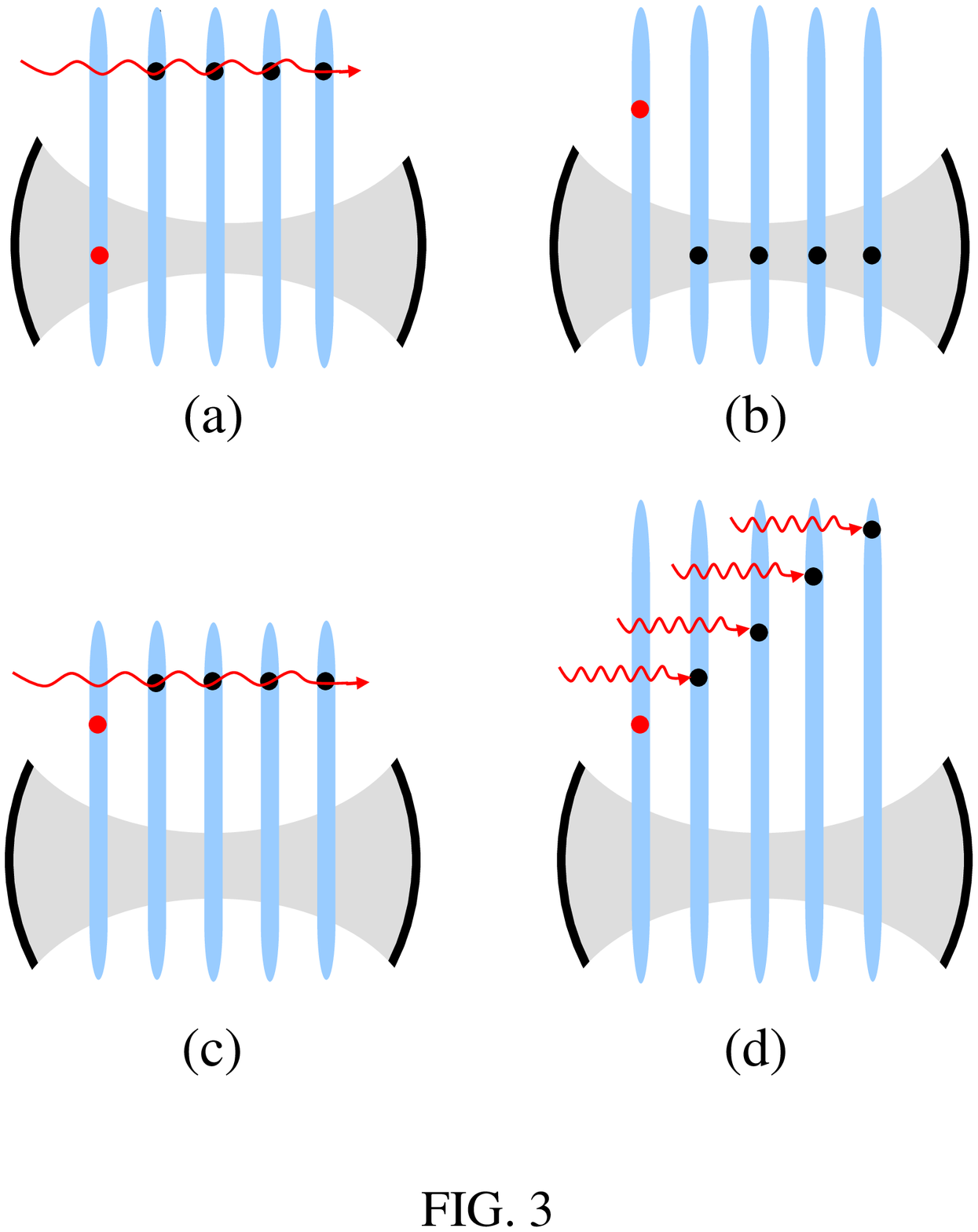} %
\vspace*{-0.08in}
\caption{(Color online) Proposed setup for the gate realization with the
control atom (the red dot), the $n$ identical target atoms (the dark dots),
and a cavity. For simplicity, only five atoms are drawn here. Each atom can
be trapped and loaded into or out of the cavity by a one-dimensional
translating optical lattice trap [11,12].}
\label{fig:3}
\end{figure}

The cavity mode is initially in the vacuum state $\left| 0\right\rangle _c.$
The procedure for realizing the gate is as follows:

\textit{Step (i)}: Apply a pulse of $\{\omega _{21},$ $-\pi /2,$ $\pi
/(4\Omega _r)\}$ to atoms ($2,3,...,n+1$), to transform the state $\left|
1\right\rangle _k$ of atom $k$ to $(\left| 1\right\rangle _k+\left|
2\right\rangle _k)/\sqrt{2}$ ($k=2,3,...,n+1$). Meanwhile, move atom $1$
into the cavity for a time $\pi /(2g_r)$ [Fig.~3(a)]$,$ to transform the
state $\left| 1\right\rangle _1\left| 0\right\rangle _c$ to $-i\left|
0\right\rangle _1\left| 1\right\rangle _c$ but leave the state $\left|
0\right\rangle _1\left| 0\right\rangle _c$ unchanged$.$ Then, move atom $1$
out of the cavity.

\textit{Step (ii):} Move atoms ($2,3,...,n+1$) into the cavity for a time $%
\pi \Delta _c/g^2$ [Fig.~3(b)]. As a result, the states $\left|
0\right\rangle _k\left| 0\right\rangle _c,$ $\left| 1\right\rangle _k\left|
0\right\rangle _c,$ $\left| 2\right\rangle _k\left| 0\right\rangle _c,\left|
0\right\rangle _k\left| 1\right\rangle _c,$ and $\left| 1\right\rangle
_k\left| 1\right\rangle _c$ remain unchanged; but the state $\left|
2\right\rangle _k\left| 1\right\rangle _c$ changes to $-\left|
2\right\rangle _k\left| 1\right\rangle _c$ ($k=2,3,...,n+2$) as discussed
previously. Then, move atoms ($2,3,...,n+1$) out of the cavity.

\textit{Step (iii):} Apply a pulse of $\{\omega _{21},$ $\pi /2,$ $\pi
/\left( 4\Omega _r\right) \}$ to atoms ($2,3,...,n+1$), to transform the
state $(\left| 1\right\rangle _k+\left| 2\right\rangle _k)/\sqrt{2}$ of atom
$k$ to $\left| 1\right\rangle _k$ while the state $(\left| 1\right\rangle
_k-\left| 2\right\rangle _k)/\sqrt{2}$ to $-\left| 2\right\rangle _k$ ($%
k=2,3,...,n+1$).

\textit{Step (iv):} Adjust the positions of atoms ($2,3,...,n+1$) (e.g., by
translating optical lattices) such that the atoms are sufficiently separated
in space and then apply a classical pulse to each of them [Fig.~3(d)]. Each
pulse has the same frequency and a zero initial phase. The pulse applied to
atom $k$ ($k=2,3,...,n+1$) is off-resonant with the $\left| 2\right\rangle
_k\leftrightarrow \left| 3\right\rangle _k$ transition of atom $k$ with a
detuning $\Delta =\omega _{32}-\omega ,$ which has a Rabi frequency $\Omega
_k$ and a duration $\tau .$ As discussed previously, a phase shift $\theta
_k=\Omega _k^2\tau \Delta $ on the state $\left| 2\right\rangle $ of atom $k$
is obtained after the pulse is applied to atom $k.$ Note that $\theta _k$
can be adjusted by changing the pulse Rabi frequency $\Omega _k.$

\textit{Step (v)}: Adjust the positions of the atoms ($2,3,...,n+1$) back to
the original positions as depicted in Fig.~3(c) and then apply a pulse of $%
\{\omega _{21},$ $-\pi /2,$ $\pi /\left( 4\Omega _r\right) \}$ to atoms ($%
2,3,...,n+1$), to transform the state $\left| 1\right\rangle _k$ of atom $k$
to the state $(\left| 1\right\rangle _k+\left| 2\right\rangle _k)/\sqrt{2}$
while the state $\left| 2\right\rangle _k$ to $-(\left| 1\right\rangle
_k-\left| 2\right\rangle _k)/\sqrt{2}$ ($k=2,3,...,n+1$) $.$

\textit{Step (vi): }Move atoms ($2,3,...,n+1$) back into the cavity for a
time $\pi \Delta _c/g^2$ [Fig.~3(b)]. The results for this step of operation
are the same as those given in step~(ii). Then, move the atoms ($2,3,...,n+1$%
) out of the cavity.

\textit{Step (vii):} Apply a pulse of $\{\omega _{21},$ $\pi /2,$ $\pi
/(4\Omega _r)\}$ to atoms ($2,3,...,n+1$) [Fig.~3(a)], to transform the
state $(\left| 1\right\rangle _k+\left| 2\right\rangle _k)/\sqrt{2}$ of atom
$k$ to $\left| 1\right\rangle _k$ ($k=2,3,...,n+1$). Meanwhile, move atom $1$
back into the cavity for a time $3\pi /(2g_r)$ [Fig. 3(a)]$,$ such that the
state $\left| 0\right\rangle _1\left| 1\right\rangle _c$ changes to $i\left|
1\right\rangle _1\left| 0\right\rangle _c$ while the state $\left|
0\right\rangle _1\left| 0\right\rangle _c$ remains unchanged$.$ Then, move
atom $1$ out of the cavity.

One can check that the $(n+1)$-qubit phase gate depicted in Fig. 1(a) was
obtained with atoms (i.e., the control atom 1 and the target atoms $2,3,...,$
and $n+1$) after the above manipulation.

Simultaneous interaction of a pulse with all of the target atoms ($%
2,3,...,n+1$) during steps (i), (iii), (v), and (vii) is unnecessary.
Instead, one can apply a pulse to each or part of the atoms ($2,3,...,n+1$)
separately.

For step~(iv), one can also set $\Omega _2=$ $\Omega _3=...=\Omega
_{n+1}=\Omega .$ In this case, $\theta _k=\Omega ^2\tau _k\Delta ,$
which can be tuned by changing the duration $\tau _k$ of the pulse applied
to atom $k.$

\begin{figure}[tbp]
\includegraphics[bb=0 0 514 335, width=8.6 cm, height=5.0 cm, clip]{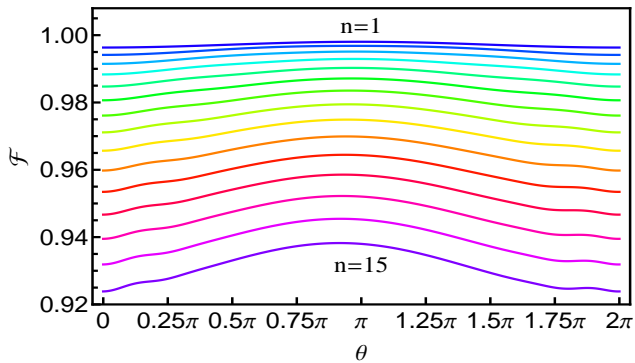} %
\vspace*{-0.08in}
\caption{(Color online) Fidelity versus $\theta $.
Lines from top to bottom correspond to $n=1$, $n=2$,..., and $n=15$,
respectively. Here, $n$ is the number of target atoms ($2,3,...,n+1$).}
\label{fig:4}
\end{figure}

Due to non-exact placement of each atom in a preselected point of the cavity
axis, the real coupling constant $g_k$ ($g_r^{\prime }$) between the target
atom $k$ (the control atom $1$) and the cavity may be different from the
ideal $g$ ($g_r$) above. Hence, it is necessary to investigate how the gate
fidelity is affected by the deviation of the real coupling constants from
the ideal ones. The fidelity is defined by $\mathcal{F}=\left| \left\langle
\psi _{id}\right| \left. \psi \right\rangle \right| ^2,$ where $\left| \psi
_{id}\right\rangle $ and $\left| \psi \right\rangle $ are the states of the
system after the above operations for the ideal case and the nonideal case,
respectively. For the latter case, the effective Hamiltonian, associated
with steps\ (ii) and (vi), takes the same form as the Hamiltonian (2) with $g
$ replaced by $g_k$. As an example, consider that each atom is initially in
the state $\left( \left| 0\right\rangle +\left| 1\right\rangle \right) /%
\sqrt{2}$ before the gate. One can easily work out the states $\left| \psi
_{id}\right\rangle $ and $\left| \psi \right\rangle $ (not shown to simplify
our presentation). We have plotted Fig. 4 to demonstrate how the fidelity
changes versus $\theta \in \left[ 0,2\pi \right] $ for $\theta _k=\theta ,$ $%
g_k=0.99g$ ($k=2,3,...,n+1$), and $\widetilde{g}_r=0.99g_r$. Fig. 4 shows
that the gate fidelity decreases as the number $n$ of target qubits
increases but a high fidelity $\sim 0.96$ or more can be achieved when $%
n\leq 10.$

\textit{Multiqubit phase gate in QFT---}For a two qubit CP gate described by
$\left| 00\right\rangle \rightarrow \left| 00\right\rangle ,\left|
01\right\rangle \rightarrow \left| 01\right\rangle ,\left| 10\right\rangle
\rightarrow \left| 10\right\rangle ,$ and $\left| 11\right\rangle
\rightarrow e^{i\varphi }\left| 11\right\rangle ,$ it is obvious that the
roles of the two qubits can be interchanged. Thus, the multiqubit phase gate
in Fig.~1(a) is equivalent to the one with $n$-control qubits (qubits $%
2,3,...,n+1$) acting on one target qubit (qubit $1$) [Fig.~1(b)]. The gate
in Fig.~1(b), with $\theta _k=$ $2\pi /2^k$ ($k=2,3,...n$), plays an
important role in quantum Fourier transforms (QFT). To implement this
gate, the pulse Rabi frequencies $\Omega _2,$ $\Omega _3,...,$ and $\Omega
_{n+1}$, involved in step (iv), need to satisfy the relation $\Omega
_{k+1}/\Omega _k=1/\sqrt{2}$ ($k=2,3,...,n$), and the operation time $\tau $
for step (iv) needs to be set by $\tau =(\Delta /\Omega _2^2)(2\pi /2^2).$
In this way, one can obtain $\theta _k=\Omega _k^2\tau /\Delta =$ $2\pi /2^k$
($k=2,3,...,n+1$).

\textit{Fidelity}---Steps (i), (iii), (v), and (vii) can be completed within
a very short time because of using resonant interactions only. Thus the
dissipation of the atoms ($2,3,...,n+1$) and the cavity for these steps is
negligibly small. One can choose atom $1$ with sufficiently long spontaneous
emission time such that decoherence of this atom is negligible during the
entire operation. Thus, the dissipation of the system
would appear in steps (ii) and (vi) due to the use of the atom-cavity or
atom-pulse dispersive interaction.

During step (ii) or step (vi), the dynamics of the lossy system, composed of
the cavity mode and atoms ($2,3,...,n+1$), is determined by
\begin{eqnarray}
\frac{d\rho }{dt} &=&-i\left[ H,\rho \right] +\kappa \mathcal{L}\left[
a\right] +\sum_{i=2,1,0}\sum_{k=2}^{n+1}\gamma _{3i}\mathcal{L}\left[ \sigma
_{i3,k}^{-}\right]   \nonumber \\
&&+\sum_{i=1,0}\sum_{k=2}^{n+1}\gamma _{2i}\mathcal{L}\left[ \sigma
_{i2,k}^{-}\right] +\sum_{k=2}^{n+1}\gamma _{10}\mathcal{L}\left[ \sigma
_{01,k}^{-}\right] ,
\end{eqnarray}
where $H$ is the Hamiltonian (1) with $g$ replaced by $g_k$,
$\mathcal{L}\left[ a\right] =\left( 2a\rho a^{+}-a^{+}a\rho -\rho
a^{+}a\right) ,$ $\mathcal{L}\left[ \sigma _{ij,k}^{-}\right] =2\sigma
_{ij,k}^{-}\rho \sigma _{ij,k}^{+}-\sigma _{ij,k}^{+}\sigma _{ij,k}^{-}\rho
-\rho \sigma _{ij,k}^{+}\sigma _{ij,k}^{-}$ (with $\sigma _{ij,k}^{-}=\left|
i\right\rangle _k\left\langle j\right| $, $\sigma _{ij,k}^{+}=\left|
j\right\rangle _k\left\langle i\right| $ and $ij\in \{23,13,03,12,02,01\}$),
$\kappa $ is the decay rate of the cavity mode, $\gamma _{ji}$ is the decay
rate of the level $\left| j\right\rangle $ of atoms ($2,3,...,n+1$) via the
decay path $\left| j\right\rangle \rightarrow \left| i\right\rangle $ (here,
$ji\in \{32,31,30,21,20,10\}$). In addition, during step (iv), the dynamics
of the lossy system, composed of the cavity mode and atoms ($2,3,...,n+1$%
), is described by $d\rho /dt=-i\left[ \sum_{k=2}^{n+1}H_I^k,\rho \right] +%
\mathcal{E},$ where $H_I^k$ is the Hamiltonian (3) and $\mathcal{E}$ is the
sum of the last four terms of Eq.~(4).

The fidelity of the gate operations is given by $\mathcal{F}=\left\langle
\psi _{id}\right| \widetilde{\rho }\left| \psi _{id}\right\rangle ,$ where $%
\left| \psi _{id}\right\rangle $ is the state of the whole system after the
gate operations, in the ideal case without considering the dissipation of
the system during the entire operation and non-exact placement of atoms
inside the cavity; and $\widetilde{\rho }$ is the final density operator of
the whole system after the gate operations are performed in a real situation.

\begin{figure}[tbp]
\begin{center}
\includegraphics[bb=2 4 453 293, width=8.6 cm, height=4.0 cm, clip]{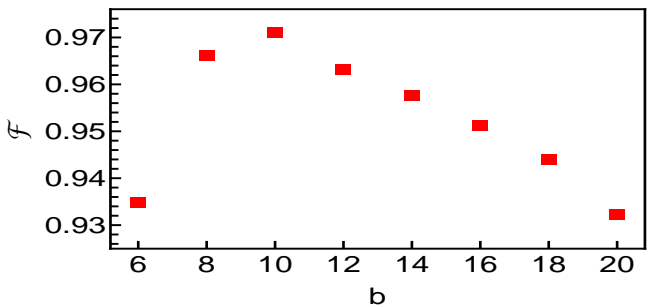} %
\vspace*{-0.08in}
\end{center}
\caption{(Color online) Fidelity versus $b$. Here, $b=\Delta _c/g=\Delta
/\Omega _2.$}
\label{fig:5}
\end{figure}

As an example, let us consider realizing a four-qubit CP gate in QFT, using
a two-level atom $1$ and three identical four-level atoms ($2,3,4$). The
three identical atoms ($2,3,4$) are chosen as Rydberg atoms with the
principle quantum numbers 49, 50, and 51, which correspond to the three
levels $\left| 1\right\rangle ,$ $\left| 2\right\rangle ,$ and $\left|
3\right\rangle $ as depicted in Fig. 2, respectively. We label the energy
relaxation times for the three levels $\left| 1\right\rangle ,$ $\left|
2\right\rangle ,$ and $\left| 3\right\rangle $ by $\gamma _1^{-1},\gamma
_2^{-1},$ and $\gamma _3^{-1},$ which are on the order of $\sim 3\times
10^{-2}$ s (e.g., see [13-15]). Without loss of generality, assume that each
of the four atoms is initially in the state $\left( \left| 0\right\rangle
+\left| 1\right\rangle \right) /\sqrt{2}$ and the cavity mode is in the
vacuum state before the gate. The expression for the ideal state $\left|
\psi _{id}\right\rangle $ of the system after the entire operation is
straightforward (not shown here to simply our presentation). As a
conservative estimation, consider $\gamma _{32}^{-1}=\gamma
_{31}^{-1}=\gamma _{30}^{-1}=\gamma _3^{-1},$ $\gamma _{21}^{-1}=\gamma
_{20}^{-1}=\gamma _2^{-1},$ and $\gamma _{10}^{-1}=\gamma _1^{-1}$. In
addition, choose $\widetilde{g}_r=0.99g_r$, $g_2=g_3=g_4=0.99g$, $g_r=g=2\pi
\times 50$ KHz [13,14], $\Omega _r=g_r,$ $\Omega _2=g,$ $\Omega _4/\Omega
_3=\Omega _3/\Omega _2=1/\sqrt{2},\tau =\left( \Delta /\Omega _2^2\right)
\left( 2\pi /2^2\right) ,\tau _m=1$ $\mu $s (a typical time for loading
atoms into or out of the cavity), and $\kappa ^{-1}=3.0\times 10^{-2}$ s.
Our numerical calculation shows that a high fidelity $\sim 97\%$ can be
achieved when the ratio $b=\Delta _c/g=\Delta /\Omega _2$ is about 10
(Fig.~5).

For Rydberg atoms chosen here, the $\left| 2\right\rangle \leftrightarrow
\left| 3\right\rangle $ transition frequency is $\sim 51$ GHz [15]. The
cavity mode frequency is then $\sim 50.9995$ GHz for $\Delta _c/g=10$. For
the cavity-photon lifetime $\kappa ^{-1}$ used in our calculation, the
required quality factor $Q$ of the cavity is $%
\sim 9.6\times 10^9.$ Note that cavities with a high $Q\sim 3\times 10^{10}$
was previously reported [16].

\textit{Discussion}---The idea of coupling qubits to a cavity to implement
the proposed gate was previously presented [17]. However, our present
proposal differs from the one in [17]. First, in our present proposal,
application of pulses to the qubits is carried out \textit{outside} a
cavity, while the one in [17] requires that each pulse was applied to a
different target qubit \textit{inside} a cavity. Second, the present
proposal is based on a first-order large detuning, i.e., $\Delta _c\gg g$ or
$\Delta \gg \Omega _k$, while the proposal in [17] was based on a
second-order large detuning $\delta =\Delta _c-\Delta \gg g^2/\Delta
_c,\Omega _k^2/\Delta $ between the cavity mode and the pulse (for the
detailed discussion, see [17]).

\textit{Acknowledgments.}---This work was supported in part by the National
Natural Science Foundation of China under Grant Nos. 11074062, 11147186,
11174081 and 11034002, the National Basic Research Program of China under
Grant No. 2011CB921602, the Zhejiang Natural Science Foundation under Grant
No. Y6100098, and the Open Fund from the SKLPS of ECNU.

\end{document}